\documentclass[aps,prb,reprint,superscriptaddress,nobibnotes]{revtex4-2}

\usepackage{graphicx}
\usepackage{dcolumn}
\usepackage{bm}
\usepackage{lineno}
\usepackage{ulem}
\usepackage{soul}
\usepackage{xcolor}
\usepackage{array}
\usepackage{booktabs}
\usepackage{multirow, makecell}
\usepackage{physics}
\usepackage{amssymb}
\usepackage[colorlinks=true, allcolors=blue]{hyperref}%

\newcommand{\CeIrBr}{CeIr$_{3}$B$_{2}$}

\begin{document}


\title{One-dimensionality signature in optical conductivity of heavy-fermion \CeIrBr}


\author{Bo Gyu Jang}
\affiliation{Theoretical Division, Los Alamos National Laboratory, Los Alamos, New Mexico 87545, USA}

\author{Kenneth R. O'Neal}
\affiliation{Center for Integrated Nanotechnologies, Los Alamos National Laboratory, Los Alamos, New Mexico 87545, USA}

\author{Christopher Lane}
\affiliation{Theoretical Division, Los Alamos National Laboratory, Los Alamos, New Mexico 87545, USA}
\affiliation{Center for Integrated Nanotechnologies, Los Alamos National Laboratory, Los Alamos, New Mexico 87545, USA}

\author{Thomas U. B\"ohm}
\affiliation{Institute for Materials Science, Los Alamos National Laboratory, Los Alamos, New Mexico 87545, USA}

\author{Nicholas Sirica}
\affiliation{Center for Integrated Nanotechnologies, Los Alamos National Laboratory, Los Alamos, New Mexico 87545, USA}

\author{Dmitry Yarotski}
\affiliation{Center for Integrated Nanotechnologies, Los Alamos National Laboratory, Los Alamos, New Mexico 87545, USA}

\author{Eric D. Bauer}
\affiliation{Materials Physics and Application Division, Los Alamos National Laboratory, Los Alamos, New Mexico 87545, USA}

\author{Filip Ronning}
\affiliation{Institute for Materials Science, Los Alamos National Laboratory, Los Alamos, New Mexico 87545, USA}

\author{Rohit Prasankumar}
\affiliation{Center for Integrated Nanotechnologies, Los Alamos National Laboratory, Los Alamos, New Mexico 87545, USA}

\author{Jian-Xin Zhu}
\email{jxzhu@lanl.gov}
\affiliation{Theoretical Division, Los Alamos National Laboratory, Los Alamos, New Mexico 87545, USA}
\affiliation{Center for Integrated Nanotechnologies, Los Alamos National Laboratory, Los Alamos, New Mexico 87545, USA}


\date{\today}

\begin{abstract}

In low dimensions, the combined effects of interactions and quantum fluctuations can lead to  dramatically new physics distinct from that existing in higher dimensions.
Here, we investigate the electronic and optical properties of \CeIrBr, a quasi-one-dimensional (1D) Kondo lattice system, using $ab\ initio$ calculations. The Ce atoms in the hexagonal crystal structure form 1D chains along the $c$-axis, with extremely short Ce-Ce distances. The quasi-1D nature of the crystal structure is well reflected in its electronic structure. Extremely flat bands emerge within the $ab$-plane of the Brillouin zone, yielding sharp optical transitions in the corresponding optical conductivity. Our calculations indicate that these prominent peaks in the optical conductivity provide a clear signature of quasi-1D heavy fermion systems.

\end{abstract}


\maketitle

\section{Introduction}

Heavy-fermion systems often exhibit a rich phase diagram, including magnetism, non-Fermi liquid physics, and unconventional superconductivity. The ground state of these materials is mainly governed by two major competing energy scales, the Ruderman-Kittel-Kasuya-Yoshida (RKKY) interaction and the Kondo interaction. A quantum phase transition occurs when the long-range RKKY interaction and the onsite Kondo interaction cancel each other, as one tunes external parameters such as pressure or magnetic field. If the quantum fluctuation driven transition is continuous, a quantum critical point (QCP) can be realized. In the vicinity of a QCP, exotic phenomena can be observed such as non-Fermi liquid behavior and superconductivity. 
 
Low-dimensionality enriches the anomalous behaviors of heavy-fermion systems and often modifies the critical behavior near the QCP. 
Near the pressure-induced quantum phase transition in three-dimensional (3D) bulk CeIn$_{3}$, the resistivity follows $\rho=\rho_{0}+AT^\alpha$, where $\alpha\sim1.6$ deviates from Fermi liquid behavior ($\alpha=2$)~\cite{Mathur1998, Knebel2001}. 
On the other hand, $T$-linear resistivity behavior is observed near the quantum phase transition induced by dimensionality tuning in CeIn$_{3}$/LaIn$_{3}$ superlattices~\cite{Shishido2010}.  
In addition to the dimensionality control in superlattices, there have been efforts to find low-dimensional heavy fermion materials.   Quasi-two-dimensional (2D) heavy-fermion systems have been intensively studied, such as YbRh$_{2}$Si$_{2}$, Ce122 (e.g., CeCu$_{2}$Si$_{2}$), Ce115 (e.g., CeCoIn$_{5}$), and Ce218 (e.g., Ce$_{2}$CoIn$_{8}$). 
The resistivity behavior of  CeCoIn$_{5}$~\cite{Malin2005} and YbRh$_{2}$Si$_{2}$~\cite{Friedemann2009, Nguyen2021} near the QCP is consistent with scattering from 2D antiferromagnetic fluctuations~\cite{Moriya2000}.
Also the exotic quantum criticality in CeCu$_{6-x}$Au$_x$ is related to strong magnetic two-dimensional quantum fluctuations~\cite{Rosch1997, Zhu2003, Zhu2007}.
1T/1H-TaSe$_{2}$ heterostructures, tri-layer twisted graphene, and transition metal
dichalcogenide (TMD) moir\'e materials have been suggested as artificial heavy-fermion systems without $f$ electrons~\cite{Vano2021, Ruan2021,Ramires2021,Kumar2022}.
Recently, CeSiI has been proposed to be an intrinsic vdW 2D heavy-fermion material~\cite{Jang2022}. 

There are also a few reports of quasi-1D heavy-fermion systems, such as YbNi$_{4}$P$_{2}$~\cite{Krellner2011, Steppke2013}, CeCo$_{2}$Ga$_{8}$~\cite{Wang2017}, and Ce$M_{3}A_{2}$ ($M$=Co, Rh, Ir / $A$ = B, Si)~\cite{Muro2007,Shigetoh2007,Kacz2008, Pikul2010, Kubota2013, Amorese2022prb, Amorese2022}. 
Among them, Ce$M_{3}$B$_{2}$ has received attention due to its abnormal properties. In this material, Ce atoms form 1D chains along the $c$-axis as shown in Fig. 1. The Ce-Ce distance along the chain is much shorter than the Hill limit ($\sim$3.5 \AA) and that of $\alpha$-Ce (3.41 \AA), which shows Pauli-like susceptibility~\cite{Kim2019}.  In addition, CeRh$_{3}$B$_{2}$ and CeIr$_{3}$B$_{2}$ exhibit ferromagnetic ordering, while most Ce-based heavy-fermion materials order antiferromagnetically. CeRh$_{3}$B$_{2}$ has the highest magnetic ordering temperature among Ce-based materials ($T_{C}$ = 115 K) and is followed by \CeIrBr\ ($T_{C}$ = 41 K)~\cite{Kubota2013, Amorese2022}. 

In this study,  we elucidate the electronic and optical properties of the quasi-1D heavy-fermion material, \CeIrBr ~ using $ab\ initio$ calculations. Due to the quasi-1D nature of the crystal structure, extremely flat band features occur in the $xy$-plane. The resulting optical transitions between these flat bands yield notable peak structures in the optical conductivity, which are not observed in other heavy fermion materials with higher dimensionality.

\section{Computational method}

Density functional theory (DFT) calculations were performed using the WIEN2k code, which uses a full potential linearized augmented plane-wave+local orbitals (L/APW+lo) method~\cite{Blaha2020}.  The Perdew-Burke-Ernzerhof generalized gradient approximation (PBE-GGA) was employed for the exchange-correlation potential~\cite{Perdew1996}. Spin-orbit coupling (SOC) was considered to describe the relativistic effect of heavy Ce and Ir atoms. A $10\times10\times16$ $k$-point mesh was used for self-consistent calculation.

To study the correlation effect of Ce 4$f$ electrons, we employed dynamical mean-field theory calculations combined with DFT (DFT+DMFT) as implemented in DFT+Embedded DMFT (eDMFT) functional code~\cite{Haule2010}. A hybridization window from $-10$ eV to 10 eV with respect to the Fermi level ($E_{F}$) was used, along with the Hubbard parameters  $U=4$ eV, $5$ eV, and $6$ eV and $J=0.7$ eV. The impurity model was solved using a continuous time quantum Monte Carlo (CTQMC) solver~\cite{Haule2007}.

\section{Results}

\begin{figure}
	\includegraphics{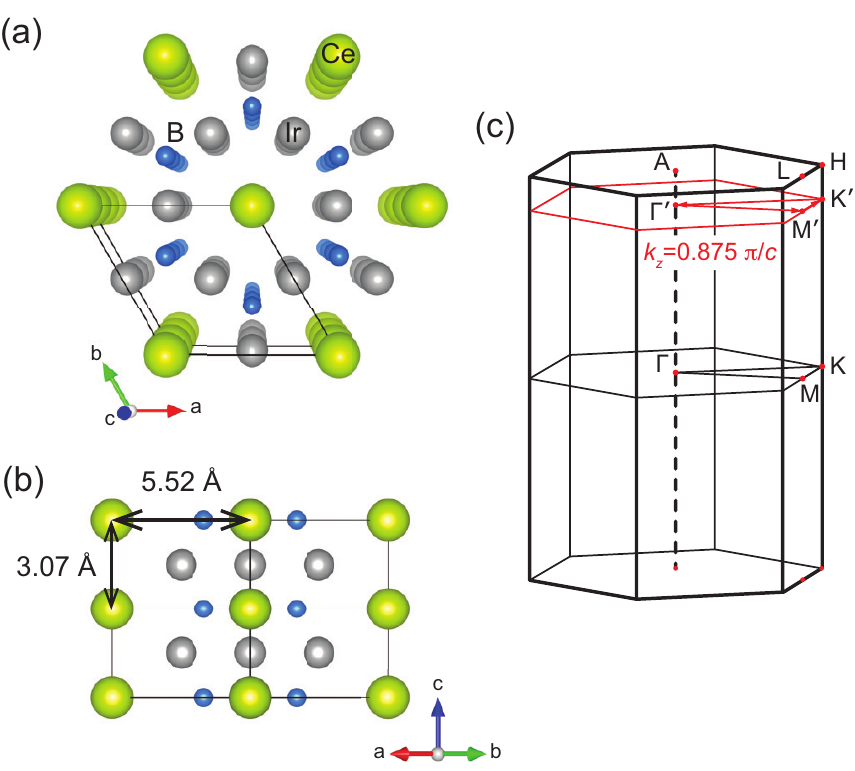}%
	\caption{Crystal structure of \CeIrBr\ is shown in (a) a top view and (b) a side view.  Green, grey, and blue indicate Ce, Ir, and B atoms, respectively. 
		           (c) The Brillouin zone of \CeIrBr.  A $k$-path on the $k_{z}=0.875\pi/c$ plane (red) is used for the band structure in Fig.~\ref{Fig3}. \label{Fig1}}
\end{figure}

Figure~\ref{Fig1} shows the crystal structure of \CeIrBr\ in the high temperature hexagonal $p6/mmm$ phase. Below 395 K, the hexagonal crystal lattice slightly distorts, precipitating a structural phase transition to a monoclinic phase~\cite{Kubota2013}. 
However, the difference between two structures is negligible so that the hexagonal structure is used throughout this study for the simplicity. In the pristine crystal, Ce atoms form quasi-1D chains along the $c$-axis, with a Ce-Ce intra-chain distance of 3.07 \AA~ and inter-chain distance of 5.52 \AA~. Ir atoms are located in the interstitial between the Ce chains facilitating weak inter-chain hopping within the $ab$-plane. Furthermore, Ir-Ir distances along the $c$ axis (3.07 \AA) and in the $ab$ plane (2.76 \AA) are comparable such that the Ir atoms form an effective 3D network, in contrast to the Ce atoms.

\begin{figure}[t]
	\includegraphics{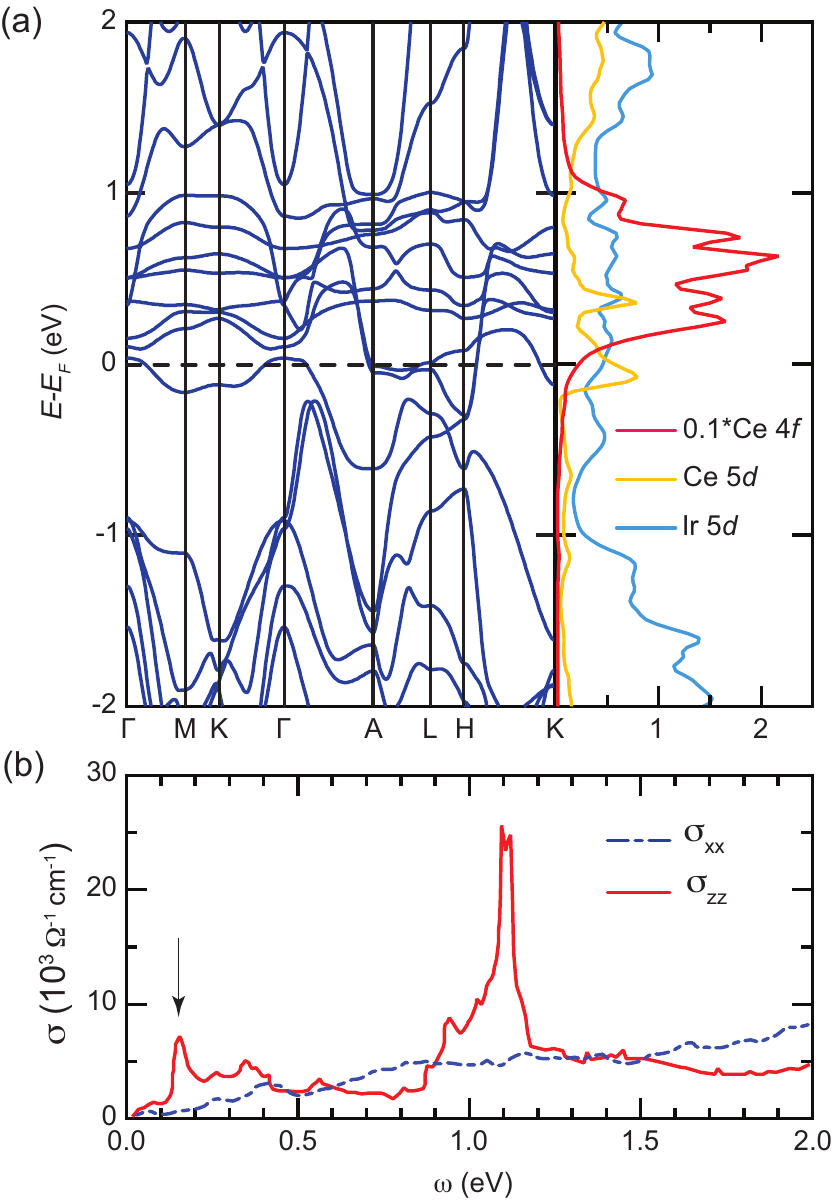}%
	\caption{(a) Electronic structure of \CeIrBr\ obtained from DFT calculations.  Red, yellow, and blue in right panel indicate Ce 4$f$, Ce 5$d$, and Ir 5$d$ partial densities of states (PDOS), respectively. Here, the Ce 4$f$ PDOS is divided by 10 for better comparison.  (b) Directional optical conductivity of \CeIrBr.  \label{Fig2}}
\end{figure}

Figure~\ref{Fig2}(a) shows the DFT electronic band structure. Bands along $k_z$ (see $\Gamma$-A and H-K paths) are more dispersive as compared to those in the plane as a result of the quasi-1D nature of the crystal structure. Concomitantly, the flat band dispersions of Ce 5$d$ orbitals within the $xy$-plane induce a sharp peak near the Fermi level ($E_{F}$) in the partial density of states (PDOS). On the other hand, the Ir 5$d$ PDOS does not display any peak structure due to the comparable bonding distances along the $c$ and in-plane directions. Moreover, due to the crystal structure geometry, Ce 4$f$ orbitals are found to strongly hybridize with Ce 5$d$ orbitals along the $z$-axis, whereas Ce-4$f$ states couple to Ir 5$d$ orbitals within the $ab$-plane.

\begin{figure*}[ht]
	\includegraphics{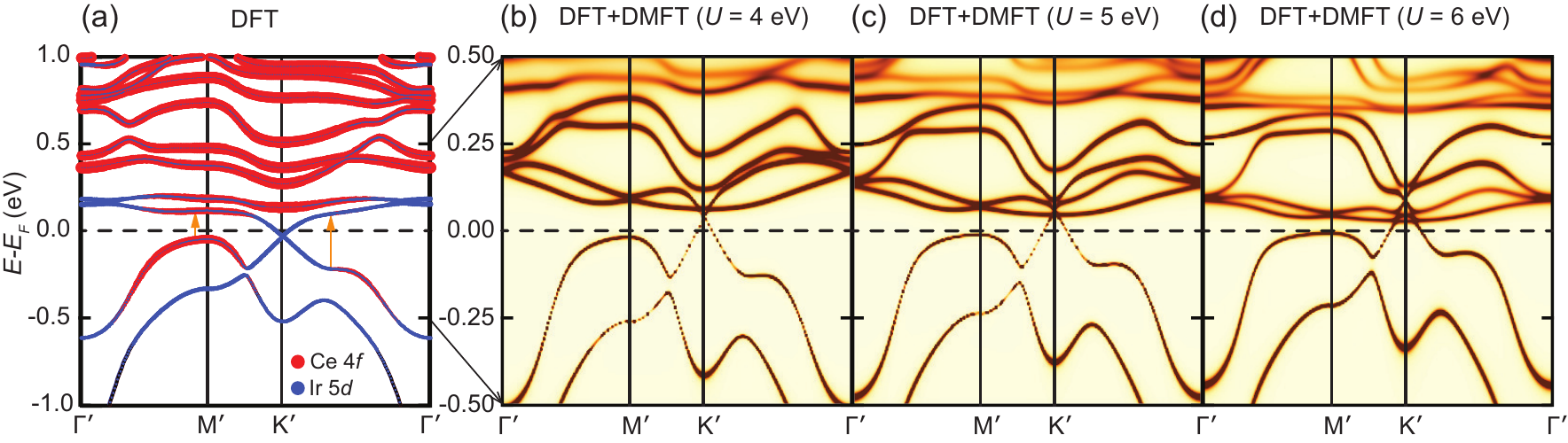}%
	\caption{Spectral function of \CeIrBr\ on the $k_{z}=0.875 \pi/c$ plane obtained from (a) DFT and (b-d) DFT+DMFT ($T$=58 K). The $k$-path used in the calculation is shown in Fig.~\ref{Fig1}(c). The possible optical transitions are shown with orange arrows in Fig.~\ref{Fig3}(a). As the $U$ value increases, the hybridization gap around the M$^\prime$ and K$^\prime$ point becomes smaller.   \label{Fig3}}
\end{figure*}

Figure~\ref{Fig2}(b) presents the associated optical conductivity obtained from the DFT bands. While the in-plane optical conductivity ($\sigma_{xx}$) monotonically increases with energy, a notable peak structure is observed along the $z$-axis ($\sigma_{zz}$). Specifically, a strong sharp peak is observed around 0.15 eV, as indicated by the black arrow. The energy scale of this optical transition is quite small,  suggesting the Ce 4$f$ orbitals participate in the optical transition.
This kind of prominent peak is not observed in the quasi-2D heavy fermion materials at low-energy scales~\cite{Singley2002, Jang2022CEF} (See also Supplementary Figure 1).

To understand the origin of this optical transition, we analyzed the momentum resolved optical matrix elements (See Supplementary Figure 2). When optical transitions are restricted to a small energy window surrounding 150 meV [120 meV - 180 meV], the non-zero matrix elements are centered on the ($\frac{1}{3}, \frac{1}{3}$, $k_{z}$) and (0,5, 0.5, $k_{z}$) momentum points in the Brillouin zone, where $k_z$ spans $-0.5$ to $0.5$. Furthermore, by scanning through the various values of $k_{z}$, the electronic bands responsible for the 150 meV optical transition are found to lie within the $k_{z}=0.875\pi/c$ plane.

Figure~\ref{Fig3} shows the band structure along the high-symmetry path in the $k_{z}=0.875\pi/c$ plane [shown in red in Fig.~\ref{Fig1}(c)]. The size of blue and red dots are proportional to the Ir 5$d$ and Ce 4$f$ orbital weight, respectively. The contribution from Ce 5$d$ orbitals was found to be negligible along this $k$-path. Extremely flat band features now clearly dominate the low-energy spectrum near the Fermi level,  thus revealing the underlying quasi-1D nature of \CeIrBr.  
The possible optical transitions are shown with orange arrows in Fig.~\ref{Fig3}(a).
Near M$^\prime$, a hybridization gap of $\sim$150 meV is observed, consistent with the sharp peak in $\sigma_{zz}$. Due to the mixture of orbital character at M$^\prime$, both Ir 5$d$ and Ce 4$f$ states are found to be involved in the optical transition. Furthermore, the weak peak at $\sim$0.35 eV in $\sigma_{zz}$ can be attributed to the transition occurring between K$^\prime$ and $\Gamma^\prime$. Overall, we find the strong sharp peaks in the optical conductivity can be regarded as a clear signature of a quasi-1D heavy-fermion system.

\begin{figure}
	\includegraphics{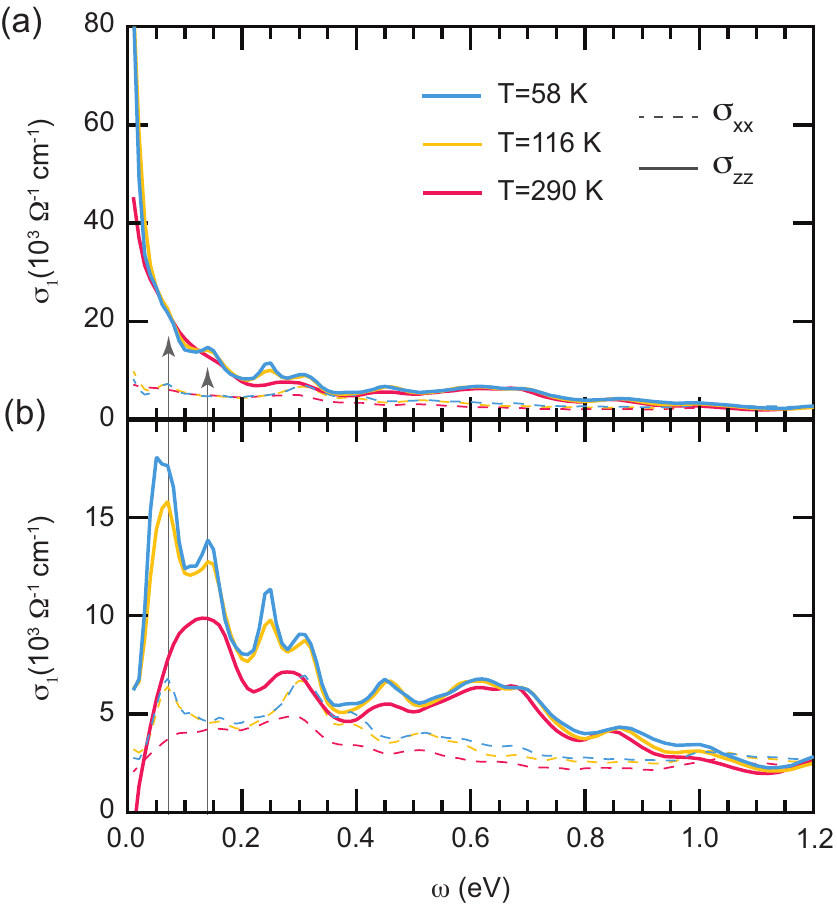}%
	\caption{Frequency-dependent optical conductivity (a) with and (b) without the Drude peak calculated within DFT+DMFT ($U=6$ eV) for varying temperature values ($T=58$ K, 116 K, and 290 K).   \label{Fig4}}
\end{figure}

Next, we performed DFT+DMFT calculations with varying Coulomb interaction $U$ values to investigate the effect of electronic correlations on the Ce 4$f$ orbitals [Fig.~\ref{Fig3} (b-d)]. Because of the strong renormalization in DFT+DMFT calculations, a small energy window is used for a better comparison to the DFT calculation in Fig.~\ref{Fig3}(a). The mass enhancement ($m^{*}/m$) estimated from the self-energy varies from 2 to 4.6 depending on $U$ value ($m^{*}/m=1-\partial$Im$\Sigma(i\omega)/\partial\omega|_{\omega \to 0^{+}}$). Not only the bandwidth of Ce 4$f$ bands but also the hybridization gap becomes smaller as $U$ value increases. The hybridization gap near the M$^\prime$ point in the $U=6$ eV calculation is $\sim75$ meV, which is almost half the size of DFT calculation. 

Figure~\ref{Fig4}(a) shows the DFT+DMFT ($U$=6 eV) total optical conductivity, including the Drude contribution, for varying temperature values ($T=58$ K, 116 K, and 290 K). The Drude peak is quite prominent in $\sigma_{zz}$, but is very weak in $\sigma_{xx}$. These features reveal clearly the quasi-1D nature of \CeIrBr. 
The anisotropic behavior of the optical conductivity is related to the anisotropic properties of effective band mass or group velocity in this system.

As the temperature increases, the Drude peak decreases in intensity. Despite this, a clear sharp Drude peak persists even up to 290 K, indicating a high Kondo temperature scale in \CeIrBr~\cite{Kubota2013, Amorese2022}. For comparison, the resistivity of CeCoIn$_{5}$ gradually increases below $\sim$200 K, which can be defined experimentally as the onset temperature of the Kondo effect~\cite{Jang2020, Lee2022}. 
Although it is difficult to define the Kondo energy scale of \CeIrBr\ from resistivity data due to the structural transition, the resistivity of \CeIrBr\ already starts to increase immediately below the structural phase transition temperature of 395 K. 
However, the short Ce-Ce distance along $c$-axis is robust against the structural phase transition.
This indicates that the extremely short Ce-Ce distance along the chain is responsible for the high Kondo temperature of this material.

Figure~\ref{Fig4}(b) shows the optical conductivity without the Drude part (only inter-band transitions). Because of the hybridization gap renormalization discussed in Fig.~\ref{Fig3}, the sharp peak observed in the DFT calculations has shifted to $\sim$75 meV. Another notable peak is observed at $\sim$0.13 eV, which corresponds to the optical transition occurring between K$^\prime$ and $\Gamma^\prime$ [See orange arrows in Fig.~\ref{Fig3}(a)]. 
Flat bands originating from the quasi-1D nature become even more flattened due to the strong mass renormalization, resulting in the enhanced optical transitions (See Supplementary Figure 3).
Although the Drude part from the intra-band transitions is quite strong, due to the short Ce-Ce distance along the $c$ direction, these notable peaks arising from inter-band transitions are still noticeable in the total optical conductivity.

There is a subtle difference between the temperature dependence of the two notable peaks. 
As the temperature increases, the peak at 75 meV diminishes more rapidly compared to the peak at 0.13 eV.
The flat valence band near M$^\prime$ moves upward and becomes incoherent, making the optical transition ill-defined (See Supplementary Figure 4).  However, the hybridization gap between K$^\prime$ and $\Gamma^\prime$ is clearly defined even at 290 K. As a result, the optical transition originating from the hybridization gap between K$^\prime$ and $\Gamma^\prime$ yields the most prominent peak at 290 K.

\section{Summary}
In summary, we have observed a prominent optical transition peak at low-energy scales in the quasi-1D heavy fermion system \CeIrBr. This kind of strong sharp peak cannot be found in heavy fermion materials with higher dimensionality. The transitions between the extremely flat bands originating from the quasi-1D Ce chains result in sharp peaks in the optical conductivity. This notable peak is robust in DFT+DMFT calculations, albeit shifted to a lower energy scale due to the strong mass renormalization. Therefore, this prominent peak feature can be regarded as a clear signature of quasi-1D heavy fermion systems.

\begin{acknowledgments}
We thank Hongchul Choi  and Chao Cao for helpful discussions.
Work at Los Alamos was carried out under the auspices of the U.S. Department of Energy (DOE) National Nuclear Security Administration
(NNSA) under Contract No. 89233218CNA000001. It was supported by the LANL LDRD Program, UC Laboratory Fees Research Program
(Grant Number: FR-20-653926), and in part by Center for Integrated Nanotechnologies, a DOE BES user facility, in partnership with the LANL Institutional Computing Program for computational resources. EDB acknowledges support from the U.S. DOE BES Materials Sciences and Engineering Division under the ''Quantum Fluctuations in Narrow Band Systems'' project.

\end{acknowledgments}

\bibliography{CeIr3B2}

\pagebreak
\newpage
\onecolumngrid
\renewcommand\figurename{Supplementary Figure}
\setcounter{figure}{0}

\begin{figure}
	\centering
	\includegraphics{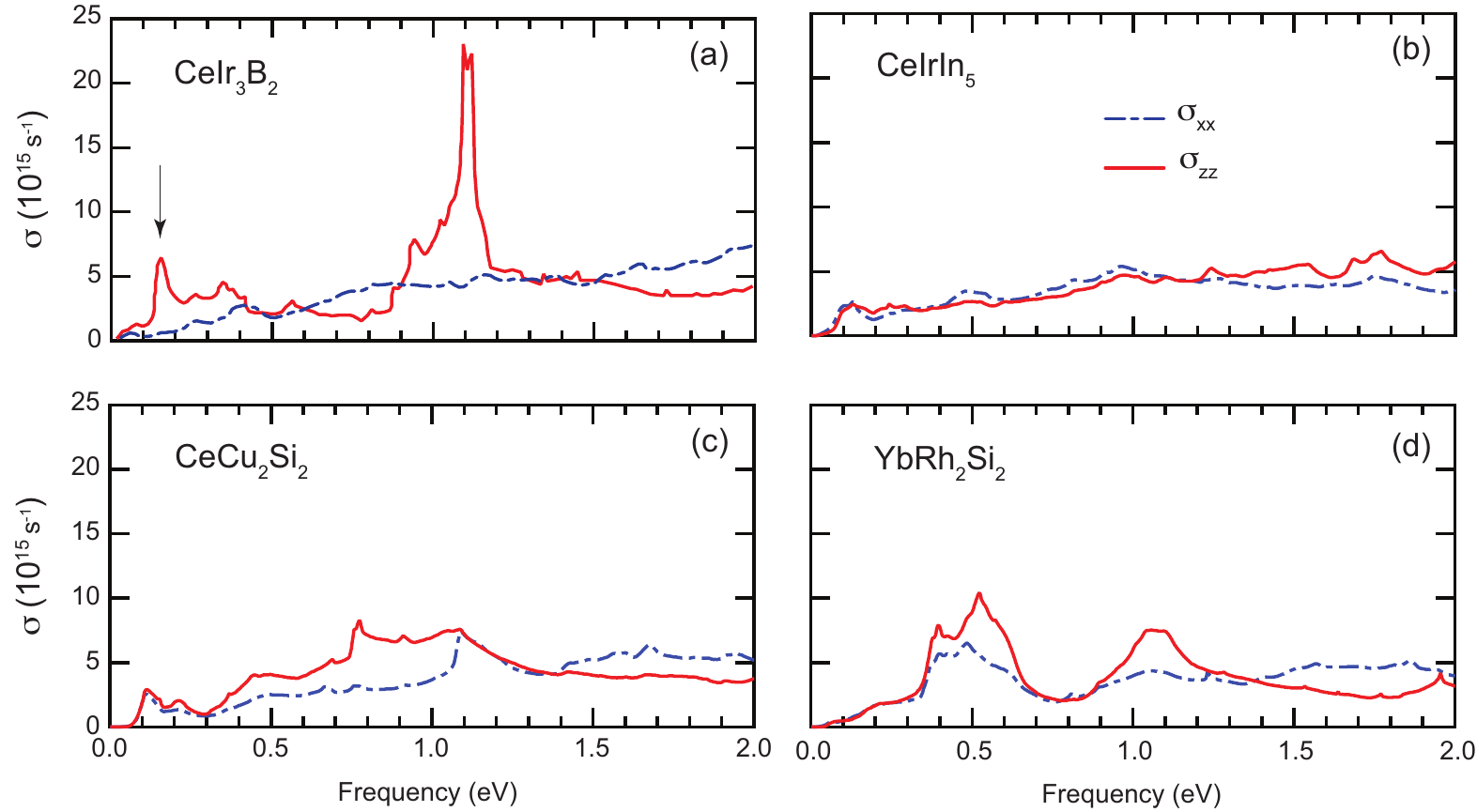}%
	\caption{Optical conductivity of  \CeIrBr\ (a) and representative quasi-2D heavy fermion materials  CeIrIn$_{5}$ (b), CeCu$_{2}$Si$_{2}$ (c), and  YbRh$_{2}$Si$_{2}$ (d). Unlike \CeIrBr, there is no notable optical transition peak at low energy in quasi-2D heavy fermion materials. Although CeCu$_{2}$Si$_{2}$ shows a peak structure at $\sim0.15$ eV, the peak insensity is much small and the peak width is also broad compared to the notable peak from \CeIrBr. }
	\label{FigS1}
\end{figure}

\begin{figure}
	\centering
	\includegraphics[width=0.8\linewidth]{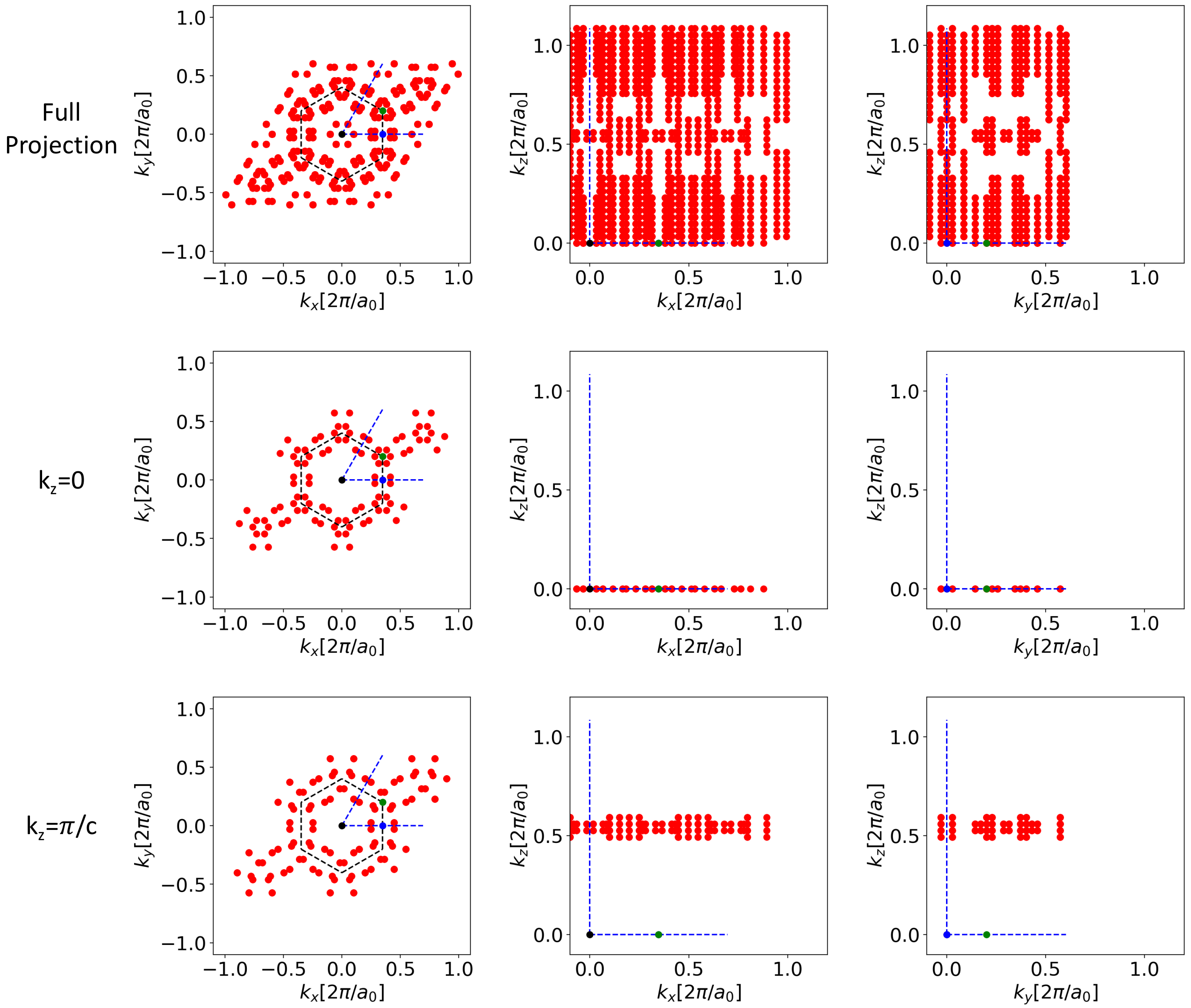}%
	\caption{Optical transition $k$-distribution in the reciprocal space. Only the optical transitions energy range between 0.12 eV and 0.18 eV are shown. Most of transitions occurs near ($k_0/2$, 0, $k_{z}$) and ($k_0$/3, $k_0$/3, $k_{z}$) points, where $-0.5 \pi/c \leq k_{z} \leq 0.5 \pi/c$. Here $k_0 = 2\pi/a_0$ with $a_0$ being the Bohr radius.}
	\label{FigS2}
\end{figure}

\clearpage

\begin{figure}
	\centering
	\includegraphics{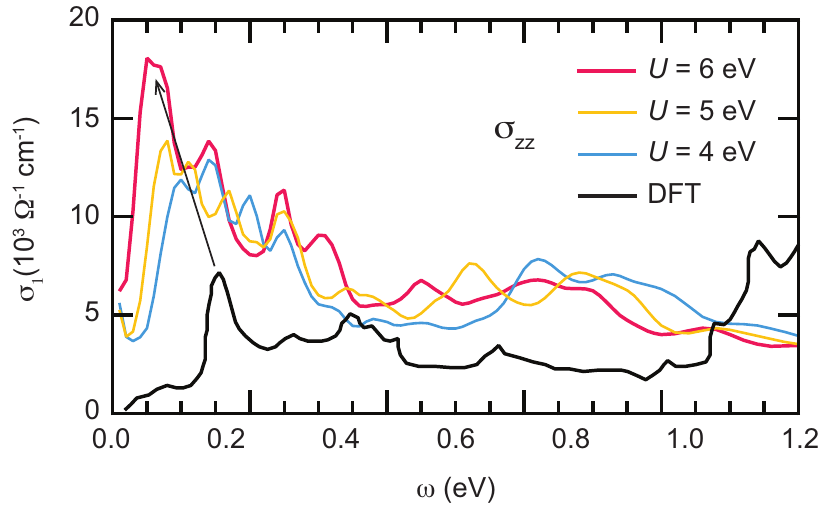}%
	\caption{Optical conductivity along the $z$-axis ($\sigma_{zz}$) obtained from DFT and DFT+DMFT with varying $U$ values ($U$=4, 5, and 6 eV). As $U$ increases, bands become flattened due to the mass renormalization. As a result, the intensity of the notable peak increases due to the enhanced optical transition between the flat bands.}
	\label{FigS3}
\end{figure}

\begin{figure}
	\centering
	\includegraphics{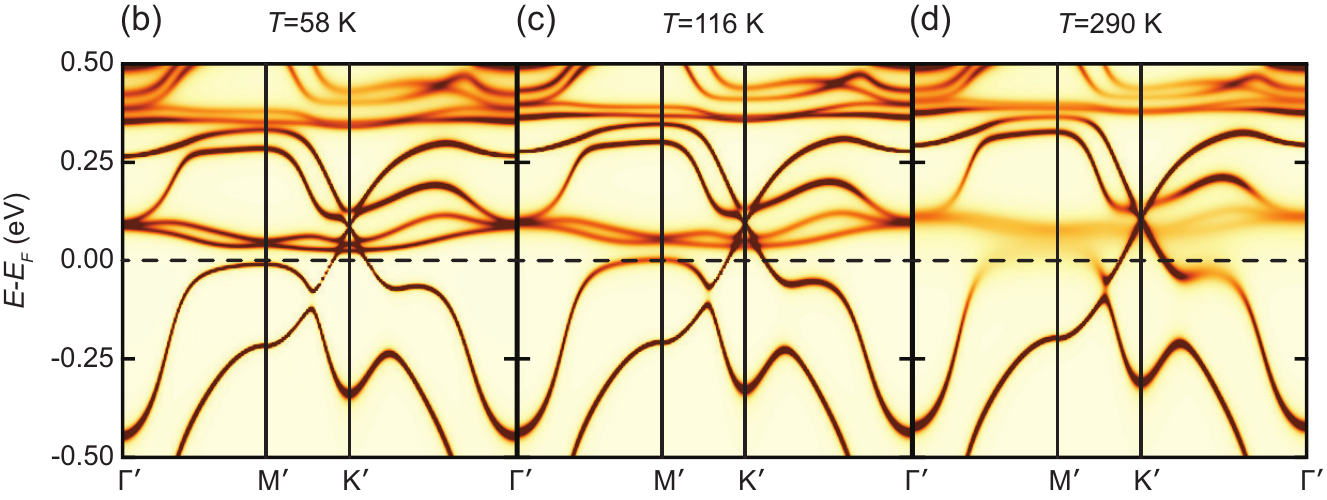}%
	\caption{Single-particle spectral function obtained from DFT+DMFT ($U=6$ eV) for temperature at $T$=58 K (a), 116 K (b), and 280 K (c). The flat valence band near M$^\prime$ point moves upward and becomes incoherent as the temperature increases.}
	\label{FigS4}
\end{figure}

\end{document}